\documentclass[a4paper,11pt]{article}
\usepackage{jheppub}

\makeatletter
\def\@fpheader{~}
\makeatother

\usepackage{bm}
\usepackage{enumerate}
\usepackage{graphicx}
\usepackage{subfigure}
\usepackage{braket}

\usepackage{array}
\usepackage{setspace}

\usepackage{hyperref}

\usepackage{amsmath,amssymb,amsfonts,amsthm,latexsym,stmaryrd}

\begin{document}

\title{Entanglement entropy and $T \overline{T}$ deformation}

\author{William Donnelly and Vasudev Shyam}
\emailAdd{wdonnelly@perimeterinstitute.ca}
\emailAdd{vshyam@perimeterinstitute.ca}
\affiliation{Perimeter Institute for Theoretical Physics, 31 Caroline St. N, N2L 2Y5, Waterloo ON, Canada}

\abstract{
Quantum gravity in a finite region of spacetime is conjectured to be dual to a conformal field theory deformed by the irrelevant operator $T \overline{T}$.
We test this conjecture with entanglement entropy, which is sensitive to ultraviolet physics on the boundary while also probing the bulk geometry. 
We find that the entanglement entropy for an entangling surface consisting of two antipodal points on a sphere is finite and precisely matches the Ryu-Takayanagi formula applied to a finite region consistent with the conjecture of McGough, Mezei and Verlinde.
We also consider a one-parameter family of conical entropies, which are finite and verify a conjecture due to Dong.
Since ultraviolet divergences are local, we conclude that the $T \overline{T}$ deformation acts as an ultraviolet cutoff on the entanglement entropy.
Our results support the conjecture that the $T \overline{T}$-deformed CFT is the holographic dual of a finite region of spacetime.
}

\maketitle

\section{Introduction}

The $T \overline{T}$ deformation of two dimensional conformal field theories (CFT) provides an exactly solvable model of quantum field theory with an ultraviolet (UV) cutoff \cite{Zamolodchikov:2004ce}.
$T \overline T$ is a composite operator satisfying the factorization property
\begin{equation} \label{factorization}
    \langle T\overline{T}\rangle=\frac{1}{8}\left(\langle T^{ab}\rangle \langle T_{ab} \rangle -\langle T^a_a \rangle^{2}\right).
\end{equation}
Any CFT can be deformed by this operator, defining a one-parameter family of theories labelled by a deformation parameter $\mu$ with dimensions of length squared.
Here we take $\mu$ to be positive.
Eq.~\eqref{factorization} is reminiscent of the factorization of multi-trace operators into products of single-trace operators in large-$N$ conformal field theories.
The factorization property determines many properties of the deformed theory, including the spectrum \cite{Zamolodchikov:2004ce} for which $\mu$ acts as a UV cutoff.
We will work in the limit of large central charge, under the assumption that $\langle T_{ab} \rangle$ can be treated as a classical field, and Eq.~\eqref{factorization} holds at all length scales.

The factorization property \eqref{factorization} determines the partition function of the $T \overline{T}$-deformed theory on an arbitrary curved background.
The response of the partition function to the change of background metric is given by the stress tensor
\begin{equation}
\delta \log Z = -\frac{1}{2} \int d^2 x \sqrt{g} \; \langle T^{a b} \rangle \delta g_{ab}.
\end{equation}
The deformed theory is defined through the flow equation:
\begin{equation} \label{Z}
\langle T^a_a \rangle = -\frac{c}{24 \pi} R - \frac{\mu}{4} \left( \langle T^{ab}\rangle \langle T_{ab} \rangle -\langle T^a_a \rangle^{2} \right)
\end{equation}
which reduces to the CFT trace anomaly in the limit $\mu \to 0$.
The flow equation together with stress tensor conservation
\begin{equation}
\nabla_a \langle T^{ab} \rangle = 0 \label{T}
\end{equation}
define a system of equations for $\langle T_{ab} \rangle$ that are in some cases sufficient to determine it completely.
In what follows we drop the angular brackets, and simply use $T_{a b}$ to refer to the expectation value.

Ref. \cite{McGough:2016lol} argued that for holographic CFTs, the $T \overline{T}$ deformation corresponds to introducing a finite boundary in the bulk acting as an infrared cutoff.
This correspondence can be most easily seen by identifying the momentum conjugate to the metric $\pi^{ab} = \sqrt{g} (T^{a b} - \tfrac{2}{\mu} g^{ab})$, under which the flow equation and stress tensor conservation become the ADM (Arnowitt-Deser-Misner \cite{Arnowitt:1962hi}) Hamiltonian constraint and momentum constraint, respectively \cite{McGough:2016lol,Shyam:2017znq}:
\begin{align}
    \frac{8\pi G}{\sqrt{g}}\left( \pi^{ab}\pi_{ab} - \left( \pi^a_a \right)^{2} \right)+\frac{\sqrt{g}}{8\pi G}\left(R+\frac{2}{\ell^{2}}\right) & =0, \\
        \nabla_{a}\pi^{ab} &= 0.
    \label{hc}
\end{align}
This fixes the identification between field theory and gravity quantities as:
\begin{equation}
c=\frac{3 \ell}{2G}, \qquad \mu = 16 \pi G \ell. \label{const}
\end{equation}
The former is the famous Brown-Henneaux central charge \cite{Brown:1986nw}, while the latter is the identification of $\mu$ proposed in \cite{McGough:2016lol}.\footnote{In \cite{McGough:2016lol} the metric of the CFT was related to the induced metric of the bulk theory by a factor $r_c$; here we set $r_c = 1$ which corresponds to measuring bulk and boundary distances in the same units.}
This proposal passes a number of consistency checks \cite{Kraus:2018xrn}.

In this letter we consider entanglement entropy of $T \overline{T}$ deformed field theory.
Entanglement entropy is a natural probe of the field theory which is sensitive to the UV induced by the $T \overline{T}$ deformation.
Moreover, it has a well-known holographic dual given by the Ryu-Takayanagi formula \cite{Ryu:2006bv}.
We will consider an entangling surface consisting of two antipodal points on the sphere, which gives the entanglement entropy of the de Sitter vacuum state across the horizon \cite{Hawking:2000da}.
We find that this entanglement entropy is finite, and matches the Ryu-Takayanagi formula with a finite cutoff surface consistent with the proposal of Ref.~\cite{McGough:2016lol}.

We further consider the conical entropy, a close relative of the R\'enyi entropy \cite{Headrick:2010zt,Hung:2011nu,Lewkowycz:2013nqa}, which carries more detailed information about the spectrum of the reduced density matrix.
These are calculated for index $n < 1$, and they verify a proposal due to Dong \cite{Dong:2016fnf} relating the conical entropy to the length of a brane anchored to the boundary. Again, the $T \overline{T}$ deformation renders the entropy UV finite, and corresponds to a finite infrared cutoff surface in the bulk.
This gives further confirmation of the conjecture of Ref.~\cite{McGough:2016lol}.

\section{Sphere partition function and entanglement entropy}

We first calculate the sphere partition function in the $T \overline{T}$-deformed CFT.
We will see that this is sufficient to calculate the entanglement entropy when the entangling surface is two antipodal points on the sphere.

To find the sphere partition function, we consider the metric $ds^2 = r^2 (d \theta^2 + \sin(\theta)^2 d \phi^2)$ and vary the radius $r$:
\begin{equation} \label{ddrlogZsphere}
\frac{d}{dr} \log Z = -\frac{1}{r} \int d^2x \sqrt{g} \; T^a_a.
\end{equation}
By symmetry, the stress tensor takes the form $T_{ab} = \alpha g_{ab}$, where $\alpha$ is determined by the flow equation \eqref{Z}:
\begin{equation} \label{alphasphere}
   \alpha = \frac{2}{\mu}\left( 1 - \sqrt{1 + \frac{c\mu}{24\pi r^2}} \right).
\end{equation}
In solving the quadratic equation we have chosen the branch that gives the CFT trace anomaly in the limit $\mu \to 0$.
This yields a differential equation for the partition function as a function of radius
\begin{equation}
    \frac{d \log Z}{dr} = \frac{16\pi}{\mu} \left( \sqrt{r^2 + \frac{c\mu}{24\pi}} - r \right).
\end{equation}
This can be integrated with the help of the substitution $r=\sqrt{\frac{c\mu}{24\pi}}\sinh(x)$, giving the sphere partition function
\begin{equation} \label{logZsphere}
\log Z = \frac{c}{3}\left(x - \frac{1}{2} e^{-2 x}\right) 
= \frac{c}{3} \sinh^{-1}\left( \sqrt \frac{24 \pi}{c \mu} r \right) + \frac{8 \pi}{\mu} \left( r \sqrt{\frac{c\mu}{24\pi} + r^2} - r^2 \right)
\end{equation}
Note that we have chosen the boundary condition $\log Z = 0$ at $r = 0$; this would not have been possible in a CFT, where the partition function continues to change as a function of scale at arbitrarily short distances.
Here we see that the flow equation is consistent with a trivial theory in the UV.

The Euclidean path integral on $S^2$ corresponds to the de Sitter vacuum.
We will consider the entanglement entropy of this state across an entangling surface consisting of two antipodal points.
This entropy can be obtained directly from the sphere partition function as follows.\footnote{We thank Aron Wall for teaching us this trick.}
To calculate the entanglement entropy by the replica trick, we consider the $n$-sheeted cover of the sphere:
\begin{equation} \label{conicalsphere}
ds^2 = r^2 (d \theta^2 + n^2 \sin(\theta)^2 d \phi^2).
\end{equation}
The entropy is then obtained from the partition function as:
\begin{equation}
S = \left(1 - n \frac{d}{dn} \right) \log Z \Big|_{n = 1}.
\end{equation}
In the absence of rotational symmetry, this formula requires analytic continuation in $n$, but in the case of antipodal points we can continuously vary $n$.

\begin{figure}
    \centering
    \includegraphics{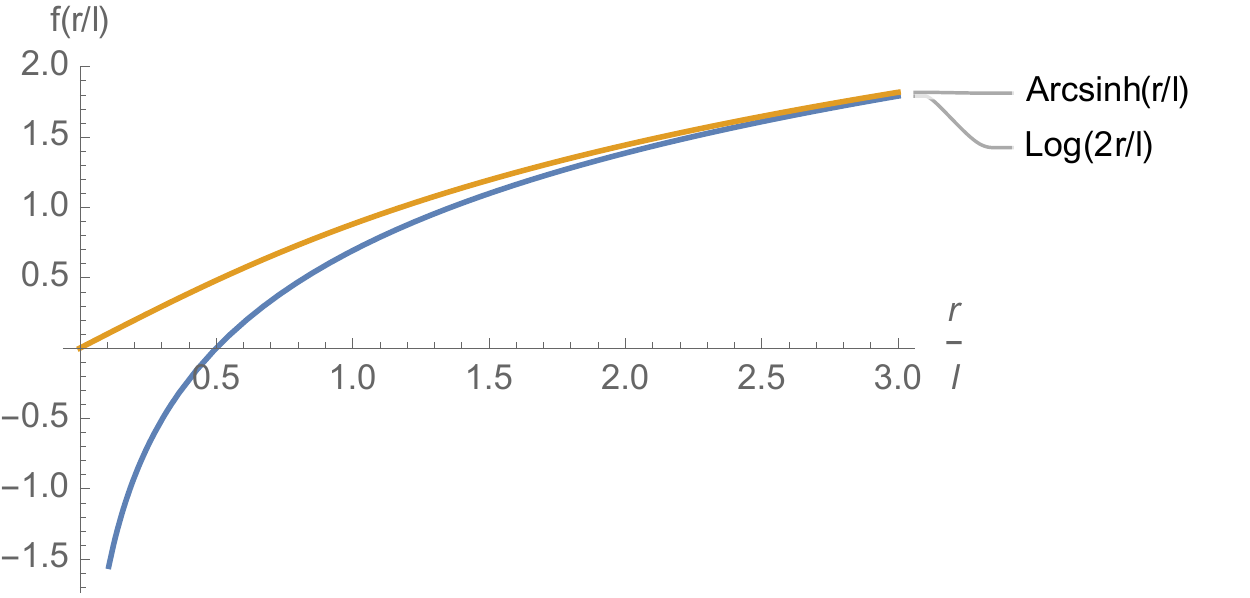}
    \caption{Entanglement entropy in the $T \overline{T}$ theory agrees with the CFT result for $r \gg \sqrt{\mu c}$, but is UV finite.}
    \label{fig:LogvsArcSinh}
\end{figure}

Under a change of $n$, the partition function changes as
\begin{equation}
    \frac{d \log Z}{dn} \Big|_{n = 1} = -\int \sqrt{g}\; T^\phi_\phi.
\end{equation}
Since the stress tensor on the sphere is isotropic, $T^\phi_\phi = \frac{1}{2}T^a_a$.
From \eqref{ddrlogZsphere} we conclude that the entropy can be expressed in terms of the sphere partition function as
\begin{equation} \label{Ssphere}
S = \left(1 - \frac{r}{2} \frac{d}{dr} \right) \log Z = \frac{c}{3} \sinh^{-1} \left(\sqrt{\frac{24\pi}{c \mu}} r \right).
\end{equation}
For $r \gg \sqrt{\mu c}$ we see that this formula reproduces the well-known CFT result \cite{Holzhey:1994we,Calabrese:2009qy} with subleading corrections (see figure \ref{fig:LogvsArcSinh}) 
\begin{equation}
S = \frac{c}{3} \log \left( \sqrt{\frac{96\pi}{c \mu}} r \right) +  \frac{c^2 \mu}{288 \pi r^{2}}  + O\left( \mu^2 \right).
\end{equation}
The corrections to the logarithmic term in the entanglement entropy are polynomial in $\mu$ starting from order one.
In the UV limit $r \to 0$ the entanglement entropy vanishes, indicating that the theory flows to a trivial theory.

We can also compare this result with the holographic proposal of \cite{McGough:2016lol}.
The metric of Euclidean AdS${}_3$ in global coordinates is
\begin{equation}
ds^2 = \ell^2 \big(d \rho^2 + \sinh(\rho)^2 (d \theta^2 + \sin(\theta)^2 d \phi^2) \big).
\end{equation}
The sphere of radius $r$ embeds into this geometry as a surface $\rho = \rho_0$, where $\rho_0$ is defined by $r = \ell \sinh(\rho_0)$.
According to the conjecture of \cite{McGough:2016lol}, the $T \overline{T}$ deformed theory is dual to quantum gravity in the region $\rho < \rho_0$. 
According to the Ryu-Takayanagi formula \cite{Ryu:2006bv}, the holographic entanglement entropy is given by $L/(4 G)$ where $L$ is the length of a minimal geodesic connecting the points of the entangling surface. In the case of antipodal points, the geodesic passes through the center and has length  $L = 2 \ell \rho_0$.
Thus the Ryu-Takayanagi formula yields:
\begin{equation}
S = \frac{L}{4 G} = \frac{\ell}{2 G} \sinh^{-1}\left( \frac{r}{\ell} \right),
\end{equation}
which agrees precisely with \eqref{Ssphere} with the identifications \eqref{const}.

\section{Conical entropy}

The entanglement entropy is just one measure of entanglement.
More information about the spectrum of the reduced density matrix is encoded in the conical entropy
\footnote{The conical entropy $\tilde S_n$ is related to the R\'enyi entropy $S_n$ as
\begin{equation}
    \tilde{S}_{n} =  n^{2}\partial_{n}\left(\frac{n-1}{n}S_{n}\right).\label{svst}
\end{equation}}:
\begin{equation} \label{Stilde}
\tilde S_n = \left( 1 - n \frac{d}{dn} \right) \log Z_n,
\end{equation}
which reduces to the entanglement entropy when $n = 1$.
In the case of two antipodal points on the sphere, $\tilde S_n$ is the de Sitter entropy at a temperature $\sim 1/n$.
By varying $n$ we probe the density of states at different energy scales.

To calculate $\tilde S_n$ via the replica trick, we consider the partition function of the theory on the conical sphere \eqref{conicalsphere}.
We will assume rotational symmetry in $\phi$, so that we can parametrize the stress tensor in terms of two nonzero components $T_{\theta \theta}(\theta)$ and $T_{\phi \phi}(\theta)$.
The problem simplifies if we consider the variables $u = \frac{\mu}{2}T^\theta_\theta - 1$ and $v = \frac{\mu}{2} T^\phi_\phi - 1$, in terms of these variables stress tensor conservation and the flow equation are
\begin{equation}
\frac{du}{d \theta} = \cot(\theta) (v - u),
\qquad uv = 1 + \frac{c\mu}{24\pi r^2}. \label{uv}
\end{equation}
This has the solution
\begin{equation} \label{cn}
u^2 = 1 + \frac{c\mu}{24\pi r^2} + \frac{c_n}{\sin(\theta)^2},
\end{equation}
where $c_n$ is independent of $\theta$.
The value of $c_n$ is determined by the coupling of the theory to the conical singularity at $\theta = 0,\pi$.

\paragraph{Boundary conditions}

We will define the theory on a surface with a conical singularity via a limiting procedure similar to Ref.~\cite{Fursaev:1995ef}.
We first define a family of smoothed replica geometries in which a small neighborhood of the conical singularity is replaced by a smooth ``cap''. 
We will take this cap to be a portion of maximally symmetric space; for $n < 1$ this is a sphere, and for $n > 1$ a hyperbolic space.

Near the conical singularity, the geometry is approximately a flat cone $ds^2 = d\tau^2 + \tau^2 n^2 d \phi^2$, from which we cut out the region $\tau < \epsilon$.
For $n > 1$ we attach this to a cap which is the region $\sigma < \sigma_c$ of the hyperbolic space with metric
$ds^2 = \ell_c^2(d \sigma^2 + \sinh(\sigma)^2 d \phi^2)$, which has constant negative curvature $R = -2/ \ell_c^2$.
Matching the intrinsic and extrinsic geometry of the circle determines
\begin{equation}
\ell_c = \frac{\epsilon n}{\sqrt{n^2-1}}, \qquad \sigma_c = \cosh^{-1}(n).
\end{equation}

The nonsingular solution on the cap takes the form $T_{ab} = \tfrac{2}{\mu}(u + 1)$ where
\begin{equation} \label{badcap}
u^2 = 1 + \frac{c\mu}{24\pi \epsilon^2} \left( \frac{1}{n^2} - 1 \right).
\end{equation}
However, \eqref{badcap} has no real solutions for small $\epsilon$ which prevents us from taking the limit $\epsilon \to 0$.\footnote{In the CFT limit this issue does not arise, because we take $\mu \to 0$ prior to taking $\epsilon \to 0$.}
In terms of bulk variables, there are no solutions when $\ell_c < \ell$: 
this corresponds to the fact that we cannot embed a hypersurface with more negative curvature than the ambient space without breaking rotational symmetry.

In the spherically symmetric case, we can instead analytically continue from $n < 1$; a similar strategy has been employed for calculating entanglement in string theory \cite{Dabholkar:1994ai,Dabholkar:1994gg,He:2014gva,Prudenziati:2018jcf}.
For $n < 1$, we can replace a neighborhood of the conical singularity with a spherical cap consisting of the region $\theta < \theta_c$ of the sphere with radius $r_c$:
\begin{equation}
    ds^2 = r_c^2 (d \theta^2 + \sin(\theta)^2 d \phi^2).
\end{equation}
Matching the length and extrinsic curvature determines
\begin{equation}
r_c = \frac{\epsilon n}{\sqrt{1 - n^2}}, \qquad \theta_c = \cos^{-1}(n).
\end{equation}
The equation for the stress tensor is given by \eqref{badcap} just as in the hyperbolic case, except that for $n < 1$ it always has real solutions.

We can now use this solution to determine the constant in \eqref{cn}.
Stress tensor conservation implies that $u$ should be continuous, which fixes the singular part of $u$:
\begin{equation} \label{u-sphere}
u^2 = 1 + \frac{c\mu}{24\pi r^2} + \frac{c\mu}{24\pi r^2 \sin(\theta)^2} \left( \frac{1}{n^2} - 1 \right).
\end{equation}
This equation determines the stress tensor 
\begin{align} \label{Tsphere}
T^\theta_\theta &= \frac{2}{\mu}\left(1 - \sqrt{1 + \frac{c\mu}{24\pi r^2} + \frac{c\mu}{24\pi r^2} \left( \frac{1}{n^2} - 1 \right) \frac{1}{\sin(\theta)^2}} \right),\\
T^\phi_\phi &= \frac{2}{\mu} \left( 1 - \frac{1 + \frac{c\mu}{24\pi r^2}}{\sqrt{1 + \frac{c\mu}{24\pi r^2} + \frac{c\mu}{24\pi r^2} \left( \frac{1}{n^2} - 1 \right) \frac{1}{\sin(\theta)^2}}} \right)
\end{align}
The sign of $u$ in \eqref{u-sphere} has been chosen so that the CFT limit $\mu \to 0$ is finite.
In this limit, the stress tensor agrees with the result obtained from the Schwarzian transformation law of the stress tensor \cite{Holzhey:1994we}.

\paragraph{Conical Entropy}

Having found an appropriate boundary condition for the stress tensor, we can now proceed to calculate the entropy $\tilde S_n$ for $n < 1$.
To find the partition function at fixed $n$, we vary the radius to obtain:
\begin{align}
\frac{d \log Z}{d r} &= - 2 \pi n r \int d \theta \sin(\theta) \left( T_\theta^\theta + T^\phi_\phi \right) \\ 
&= \frac{4\pi n}{\mu} \int d \theta \; \sin(\theta) \left[ \frac{2 r^2 + \frac{c\mu}{24\pi} + \alpha^2}{\sqrt{r^2 + \alpha^2}} - r \right]
\end{align}
where we have defined
\begin{equation}
\alpha^2 = \frac{c\mu}{24\pi} \left( 1 + \left(\frac{1}{n^2} - 1 \right) \frac{1}{\sin(\theta)^2} \right).
\end{equation}
This can be integrated in $r$ to obtain
\begin{equation}
    \log Z = \frac{n}{4 G \ell} \int d \theta \sin(\theta) \left[ \ell^2 \sinh^{-1} \left( \frac{r}{\alpha} \right) + r \sqrt{\alpha^2 + r^2} - r^2 \right]. \label{logZ}
\end{equation}
For $n = 1$, this reduces to \eqref{logZsphere}.

\begin{figure}
    \centering
    \includegraphics[width=0.5\textwidth]{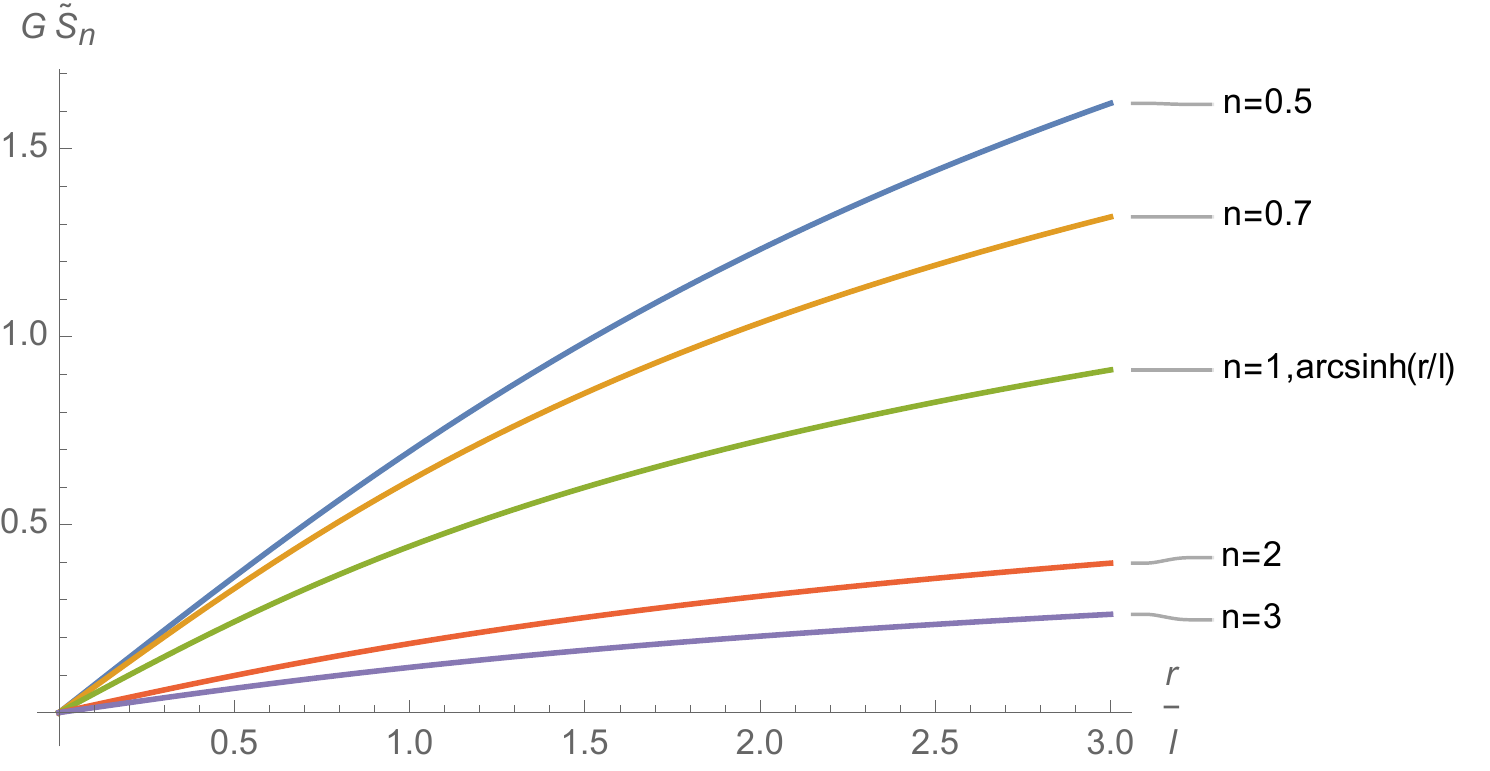}\includegraphics[width=0.5\textwidth]{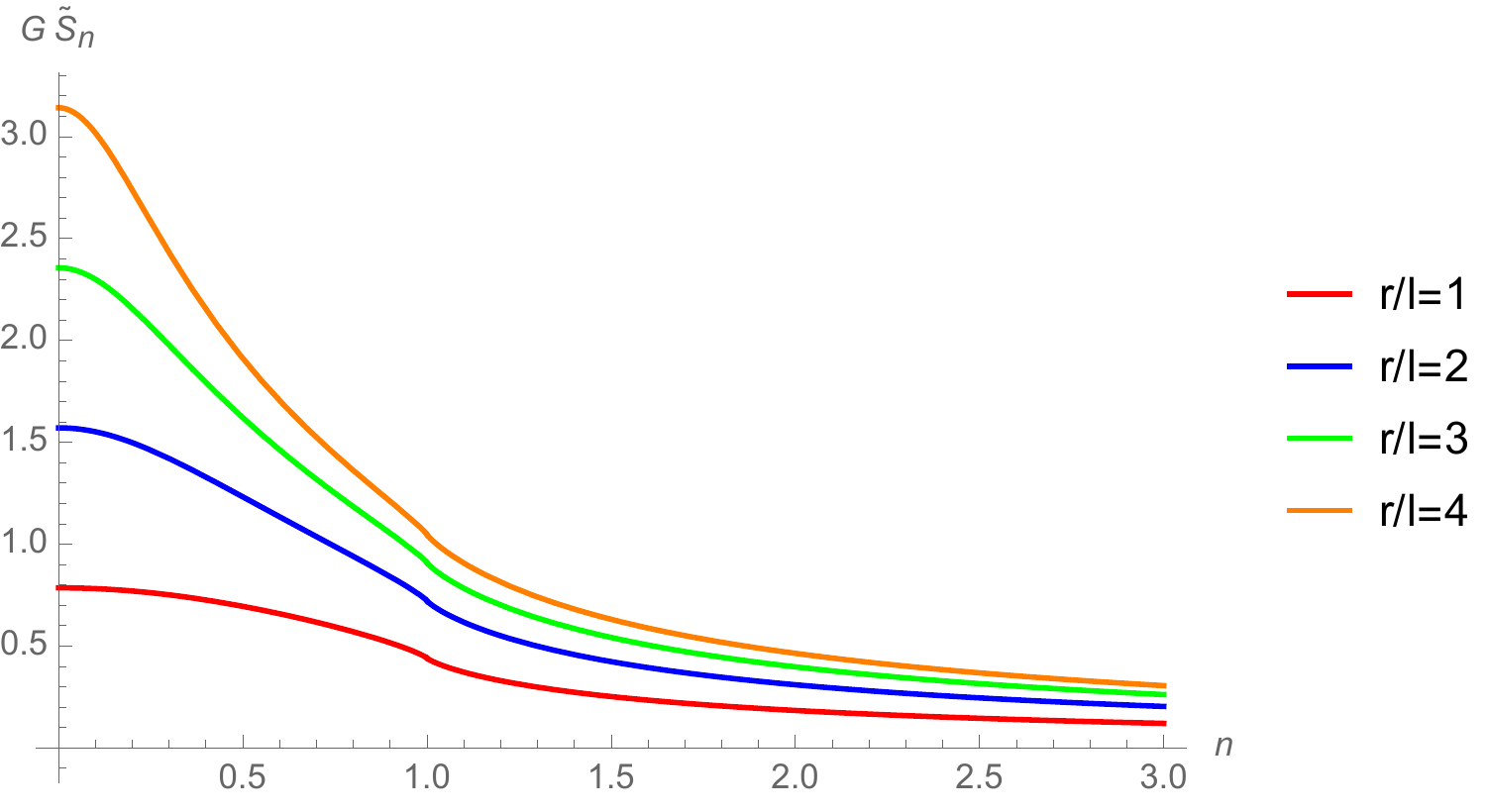}
    \caption{The entropy $\tilde S_n$ is a smoothly increasing function of $r$. 
    At fixed $r$, $\tilde S_n$ is a decreasing function of $n$, with a kink at $n = 1$.}
    \label{fig:Sr}
\end{figure}

From \eqref{logZ} we obtain the entropy
\begin{align}
    \tilde S_n &= -n^2 \frac{d}{dn} \left( \frac{\log Z}{n} \right) \\
&= \frac{r c}{6 n} \int d \theta \frac{1}{\sin(\theta)} \left( \frac{1 - \frac{c\mu}{24\pi \alpha^2}}{\sqrt{\alpha^2 + r^2}} \right) \\
&= \frac{c}{3} \frac{(1-n^2)}{\sqrt{\frac{c\mu}{24\pi r^2} + n^2}} \Pi\left(n^2 \Big| \frac{r^2 + \frac{c\mu}{24\pi}}{r^2 + \frac{c\mu}{24\pi n^2}} \right). \label{Sn}
\end{align}
$\Pi$ is the complete elliptic integral of the third kind:
\begin{equation}
    \Pi(\eta|m) := \int_0^{\pi/2} \frac{d \theta}{(1 - \eta \sin(\theta)^2) \sqrt{1 - m \sin(\theta)^2}}. \label{Pi}
\end{equation} 
This function has a branch cut for $n \geq 1$, corresponding to a pole in the integral \eqref{Pi}.
However, we can take the principal value of this integral, which is real for all $n > 0$.
This function is displayed in figure \ref{fig:Sr}. 

The limit $n \to 0$ gives the logarithm of the rank of the reduced density matrix:
\begin{equation}
\tilde S_0 = \sqrt{\frac{2 \pi c}{3 \mu}} \pi r
\end{equation}
which scales with the length, $\pi r$, of the boundary.
This is suggestive of a lattice theory in which an interval of length $L$ has a Hilbert space of dimension $\exp\left(\sqrt{\frac{2 \pi c}{3 \mu}} L \right)$.

\paragraph{Comparison with holography}

We now compare our result \eqref{Sn} with the prediction from holography.
According to the proposal of Ref.~\cite{Dong:2016fnf}, $\tilde S_n$ is given by $L/4G$, where $L$ is the length of a cosmic string whose tension induces an angle $\frac{2 \pi}{n}$ in the bulk.
By rescaling $\phi$, this is equivalent to finding the length of a geodesic in a smooth geometry with induced boundary metric given by \eqref{conicalsphere}.
Finding this solution is equivalent to finding an embedding of the metric \eqref{conicalsphere} into Euclidean AdS${}_3$.

\begin{figure}
    \centering
    \includegraphics[width=0.5\textwidth]{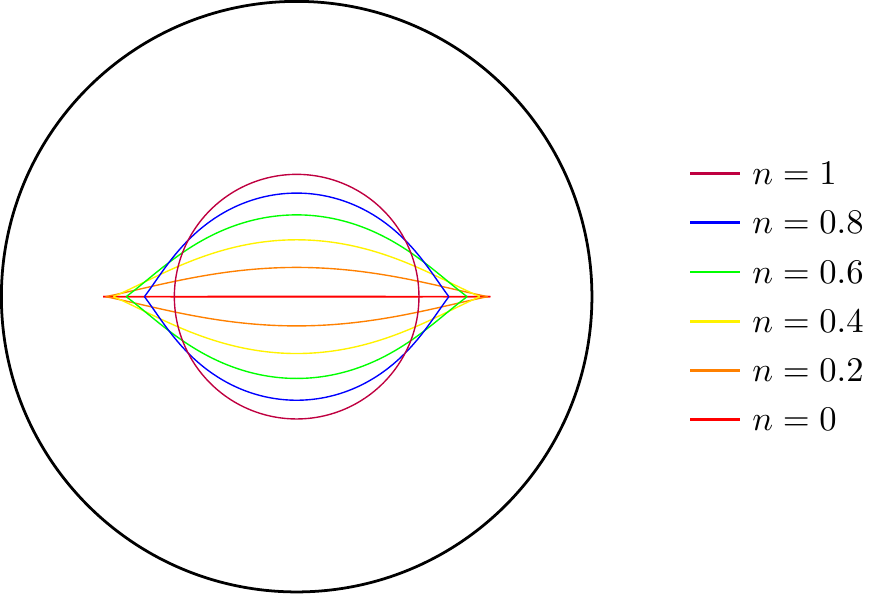}
    \caption{
    Embedded surfaces with $r = \ell = 1$ in Euclidean AdS${}_3$. 
    For this graph we have used Poincar\'e disk coordinates in which the metric is $ds^2 = \frac{4 d \vec x^2}{(1 - \vec x^2)^2}$ with the third coordinate suppressed.
    The AdS boundary is the black circle, and finite boundaries are shown in color.
    At $n = 1$, the embedding is a circle. 
    As $n$ is decreased the embeddings have an increasingly elongated ``football'' shape.
    At $n = 0$, the embedding degenerates to a line.}
    \label{fig:football}
\end{figure}

This embedding problem is most easily solved in the coordinates:
\begin{equation} \label{AdS3}
ds^2 = \ell^2 (d \varphi^2 + e^{-2 \varphi} (d\rho^2 + \rho^2 d \phi^2)).
\end{equation}
We will assume the embedding preserves rotational symmetry, so that the $\phi$ coordinate on the boundary is the same as the $\phi$ coordinate of the bulk.
The embedding is then defined by two functions $\varphi(\theta)$ and $\rho(\theta)$.
Demanding that the embedding be isometric leads to the equations
\begin{equation}
    \rho = e^\varphi \frac{r}{\ell} n \sin(\theta), \qquad \frac{r^2}{\ell^2} = \left(\frac{d \varphi}{d \theta} \right)^2 + e^{-2 \varphi} \left( \frac{d \rho}{d \theta} \right)^2.
\end{equation}
This can be rewritten as a single differential equation for 
$\varphi$,
\begin{equation}
    \frac{r^2}{\ell^2} = \left(\frac{d \varphi}{d \theta} \right)^2 + \frac{n^2 r^2}{\ell^2} \left[ \left(\frac{d \varphi}{d \theta} \right) \sin(\theta) + \cos(\theta) \right]^2,
\end{equation}
which can be solved to give
\begin{equation} \label{phiprime}
\frac{d \varphi}{d \theta} = \frac{r \sqrt{\ell^2 (1 - n^2) + n^2 (r^2 + \ell^2) \sin(\theta)^2} - n^2 r^2 \cos(\theta) \sin(\theta)}{\ell^2 + n^2 r^2 \sin(\theta)^2}.
\end{equation}
The resulting embedding in global coordinates is shown in figure \ref{fig:football}.

The length of the geodesic connecting the point $\theta = 0$ and $\theta = \pi$ in the metric \eqref{AdS3} is given by
\begin{equation}
    L = \ell (\varphi(\pi) - \varphi(0)) = \ell \int_0^\pi d \theta \, \frac{d \varphi}{d \theta}.
\end{equation}
To find the total entropy, one simply has to integrate \eqref{phiprime},
\begin{equation}
\frac{L}{4 G} = \frac{r}{2 G} \frac{1}{\sqrt{1 + r^2 n^2}} \left[ \frac{1 + r^2}{r^2} K(m) - \frac{1}{r^2} \Pi(\eta|m) \right],
\end{equation}
where $K(m) = \Pi(0|m)$ is the complete elliptic integral of the first kind and
\begin{equation}
m = \frac{n^2(1 + r^2)}{1 + n^2 r^2}, \qquad \eta = \frac{r^2 n^2}{1 + r^2 n^2}.
\end{equation}
Using an identity of elliptic integrals\footnote{$(m - \eta) \Pi(\eta|m) + (m - \eta') \Pi(\eta'|m) = m K(m)$ where $(1 - \eta) (1 - \eta') = 1 - m$ \cite[Eq. 19.7.9]{NIST:DLMF}.}, this agrees with our expression \eqref{Sn} for the entropy, with the identification of $\mu$ and $c$ given in \eqref{const}.

\section{Discussion}

We have carried out a calculation of entanglement entropy for an entangling surface consisting of two antipodal points on the sphere.
Our calculation verifies the conjecture of Ref.~\cite{McGough:2016lol} that the theory is dual to gravity in a finite region.
The entanglement entropy is UV finite and is given by the Ryu-Takayanagi formula.
Since UV divergences are local, this indicates that the entanglement entropy is finite for arbitrary entangling surfaces.

So far we have only considered the simplest case of two antipodal points, but to further test the correspondence one should consider non-antipodal points, as well as multiple intervals.
Such generalizations do not have rotational symmetry about the entangling surface, and so the method of continuation from $n < 1$ does not apply.
Instead one would have to find the appropriate solutions to equations \eqref{Z}, \eqref{T} for boundary conditions with $n > 1$.
Our analysis shows that these solutions must either be complex or break the replica symmetry.

The conical entropy $\tilde S_n$ encodes the full spectrum of the reduced density matrix.
The entanglement spectrum is given by the inverse Laplace transform of the partition function $Z$, which can be found by integrating \eqref{Stilde}.
In CFT, the $n$ dependence of $Z$ is universal, leading to a universal entanglement spectrum \cite{Calabrese:2008}.
Our result \eqref{Sn} has a highly nontrivial $n$ dependence compared to the CFT result $\tilde S_n \sim \frac{1}{n}$, suggesting a modification of the entanglement spectrum.
In particular, we would like to understand the significance of the branch cut in $\tilde S_n$, and whether the entanglement spectrum, like the spectrum on the cylinder, has imaginary eigenvalues.

The fact that the entropy vanishes as $r \rightarrow 0$ indicates that the theory is trivial in the UV. 
This flow from a trivial theory in the UV to a CFT in the infrared is reminiscent of the continuous Multiscale Entanglement Renormalization Ansatz (cMERA) \cite{Haegeman:2011uy}.
It would be interesting to see if the flow induced by $T \overline{T}$ can be viewed as locally building up entanglement as in cMERA.

The relation between the flow equation and the Hamiltonian constraint suggests that the natural generalization of $T \overline{T}$ to higher dimensions is $\frac{1}{8} (T^{ab}T_{ab} - (T^a_a)^2)$.
This generalization has been considered in the case of a flat boundary \cite{Taylor:2018xcy}, but a further generalization to curved boundaries would be required to study the entanglement entropy using the methods described here.

The $T \overline{T}$ deformation can be viewed as a result of dynamical gravity in two dimensions \cite{Freidel:2008sh}.
In particular, Dubovsky et. al. in \cite{Dubovsky:2017cnj,Dubovsky:2018bmo}, have shown that this deformation can be defined through coupling the CFT to Jackiw-Teitelboim gravity.
In a similar vein, Cardy \cite{Cardy:2018sdv} defines this deformation perturbatively through a path integral prescription involving integrating over Gaussian metric fluctuations. 
This is also consistent with the idea that putting the CFT in the bulk leads to a finite induced gravitational constant on the boundary \cite{Randall:1999vf,Emparan:2006ni,Myers:2013lva}, which leads to a finite entanglement entropy.
Similar ideas in which the entropy is regulated using an irrelevant deformation appear in Refs.~\cite{Chakraborty:2018kpr,Dong:2018cuv}.
To study the entanglement entropy from this perspective, one must face all the difficulties of defining entanglement entropy in a theory with dynamical gravity.
We hope that our results can shed some light on how gravity acts as a UV cutoff for the entanglement entropy.

\section*{Acknowledgements}

We thank Rob Myers for his comments on an earlier draft. We also thank Ed Witten and Gabriel Wong for discussions relating to this work. We would also like to thank Tomonori Ugajin for early conversations pointing us toward this interesting problem.
This research was supported in part by Perimeter Institute for Theoretical Physics. 
Research at Perimeter Institute is supported by the Government of Canada through the Department of Innovation, Science and Economic Development Canada and by the Province of Ontario through the Ministry of Research, Innovation and Science.

\bibliographystyle{utphys}
\bibliography{ttbar}

\end{document}